\documentstyle[epsfig,11pt]{article}
\begin{document}

\def\beq{\begin{eqnarray}}
\def\eeq{\end{eqnarray}}
\def\IM{{\rm {Im}}}
\def\RM{{\rm {Re}}}
\def\nn{\nonumber}
\def\Sin{{\rm sin}}
\def\Cos{{\rm cos}}
\def\PRD{Phys. Rev. D}
\def\PRL{Phys. Rev. Lett}
\def\PLB{Phys. Lett. B}
\def\EPJC{Euro. Phys. J. C}
\def\Journal#1#2#3#4{{#1}{\bf #2} (#4) #3}

\centerline{\LARGE\bf Sudakov effects in BBNS approach
    \footnote{Supported in part by National Natural Science
    Foundation of China } }

\vspace{2cm}

\centerline{ Dong-Sheng Du$\rm ^{a,b}$,
             Chao-Shang Huang$\rm ^{a,c}$,
             Zheng-Tao Wei$\rm ^{a,c}$,
             Mao-Zhi Yang$\rm ^{b}$
     \footnote{e-mail address:
      duds@mail.ihep.ac.cn;
      huangcs@itp.ac.cn;
      weizt@itp.ac.cn;
      yangmz@mail.ihep.ac.cn}}
\vspace{1cm}

\begin{center}
$\rm ^{a}$CCAST(World Laboratory), ~~P.O.Box $8730$,
Beijing $100080$, China

$\rm ^{b}$Institute of High Energy Physics, P.O.Box
$918(4)$, Beijing $100039$, China

$\rm ^{c}$Institute of Theoretical Physics, P.O.Box $2735$,
Beijing $100080$, China
\end{center}

\vspace*{0.3cm}

\begin{center} \begin{minipage}{12cm}

\noindent{\bf Abstract}

The end-point singularity is an unsolved problem in BBNS
approach. Incorporating the partonic transverse momentum and the
Sudakov form factor, this problem can be solved
model-independently. We discuss the Sudakov effects in BBNS
approach. The BBNS approach is compared with the modified PQCD
approach. The main idea of Sudakov form factor is briefly
discussed. Our conclusion is that the twist-3 contribution
for the hard spectator scattering is numerically not
important in $B\to \pi\pi$ decays, compared
with the twist-2 contribution.

\end{minipage}
\end{center}


\newpage

\section*{1.  Introduction}

The calculation of exclusive process from perturbative QCD (PQCD)
is one of the important problems in hadron physics.
Soon after the successes of PQCD application in the
deep-inelastic scattering, Drell-Yan process, etc, the
application of PQCD in some exclusive process with large momentum
transfer has been carried out and is successsful in
the asymptotic limit ($Q^2\rightarrow\infty$)~\cite{exc, Brodsky}.
The key for using PQCD is factorization, i.e., the separation
of long- and short-distance dynamics. It has been shown
in the PQCD framework for exclusive processes with large
momentum transfer that the long-distance dynamics is
involved in the light-cone hadronic wave function, the
distribution amplitude, and the physical quantity
is the convolution of the distribution amplitudes
of the initial and final hadrons  and the hard
scattering kernel~\cite{Brodsky}.

Because of the importance in exploring CP volation and
determining the CKM parameters, the exclusive, nonleptonic
two-body decays of B meson have got extensive theoretical
investigations. However, the complication caused by soft
interactions in both initial and final hadrons makes it
difficult to analyze the full QCD dynamics. Recently,
Beneke et al. proposed a QCD improved factorization
formula in exclusive B decays \cite{Beneke}, which is called
``BBNS approach" for simplicity later in this letter.
The early BSW model \cite{BSW} is the lowest order
approximation of this approach. At the order of
${\cal O}(\alpha_s)$ , the hadronic matrix element is
generally the convolution of the three light-cone distribution
amplitudes and the hard scattering kernel. For the vertex
correction and the penguin correction, it is assumed that the
$B\to M_1$ ($M_1$ denotes the meson that picks up the
spectator quark) form factor is dominated by soft interaction.
Under this assumption, the factorization formula
is simplified as the multiplicity of the $B\to M_1$ form
factor and the convolution of meson wave function with hard
scattering kernel. BBNS approach works well at leading twist level.
While for twist-3 case, it will lead to
infrared divergence in the
one-loop vertex correction and end-point singularity in hard
spectator scattering.

The problem of the infrared divergence
in the vertex correction is
partly solved in \cite{Du}. The authors used the massive gluon to
regulate the infrared divergence. They find that
the soft and collinear
divergences cancel in the vertex correction for the symmetrical
twist-3 distribution amplitude. However, the problem of end-point
singularity still remains unsolved. In \cite{Beneke2}, the
phenomenological treatment is introduced to deal with the
end-point contribution. But their treatment is artificial and
unsatisfactory. To solve this problem is the main purpose of this
paper.

The appearance of end-point singularity means that the separation
of hard and soft dynamics is not justified, so the factorization
of BBNS approach breaks down at twist-3 level. The solution of the
end-point singularity is known for a long time \cite{Brodsky}. The
method is to retain the partonic intrinsic transverse momentum
and include the mechanism of Sudakov suppression. The average of
the transverse momentum $\sqrt { <k_T^2>}\sim {\cal
O}(\Lambda_{QCD})$ in hadron is much smaller than
$M_B$. In the region far
from the end-point, the effect of transverse momentum is power
suppressed and negligible at tree level. However, in the end-point
region, the partonic transverse momentum  $k_T$ is the same order
as the longitudinal momentum $xP$, its effect is important.
Neglecting it will lead to singularity. In the hard spectator
scattering, the twist-3 contribution gives a factor $\int
\frac{du}{u^2}\phi_{\sigma}(u)$ which diverges when $u\to 0$. The
transverse momentum  smears the end-point singularity so that the
end-point contribution is not dominant. On the other hand, the
parton with transverse momentum will give rise to soft divergence
which cancels in the collinear limit. However, in the soft
region, the Sudakov suppression begins to take effect. For a
quark-antiquark pair separated by a transverse distance $b$, the
Sudakov form factor $e^{-S(Q, b)}$ suppresses the contribution
at large $b$, so that the dominant contribution comes from the
region with  small separation. This idea has been grouped into a
self-consistent and model-independent PQCD  formula. This
modified PQCD approach is given clearly in \cite{Sterman2}.
Its application in B decays can be found in \cite{Li1} and the
reference therein. As we will show, the modified PQCD approach
enlarges the range of PQCD application.

This paper is devoted to study the Sudakov effects in BBNS
approach. Here, the Sudakov effects include the transverse
momentum effects and the Sudakov form factor. We will restrict
our discussion in $B\to \pi\pi$ decays, its extension to other
$B\to PP$ decays is straightforward.  The relation of BBNS
approach and the modified PQCD approach is given  in Sect. 2. In
Sect. 3, we briefly discuss the Sudakov double logarithm and its
resummation to all orders. Some general properties of Sudakov form
factor is also disccussed. In Sect. 4, we study the Sudakov
effects in BBNS approach. The end-point singularity in the hard
spectator scattering is solved in the modified PQCD approach. In
Sect. 5, we give our conclusions and discussions.

\section*{2. BBNS approach and the modified PQCD approach }

The essential problem in $\bar B\to \pi\pi$ decay is to calculate
the hadronic matrix elements $<\pi\pi|Q_i|\bar B> $. In
\cite{Beneke}, a fatorization formula is given as: \beq
<\pi(p')\pi(q)|Q_i|\bar B(p)>= F_0^{B\pi}(q^2)\int_0^1 dv T^I
(v)\Phi_{\pi}(v)\\ \nonumber +\int_0^1 d\xi dudv
T^{II}(\xi,u,v)\Phi_B(\xi)\Phi_{\pi}(u)\Phi_{\pi}(v) \eeq
In BBNS approach, the $B\to \pi$ transition form factor $F_0^{B
\pi}$ is assumed to be dominated by soft interactions
and treated as a nonperturbative input parameter.
In this study, we hold this assumption and leave the discussion
about it in the last section.

There is only one scale $m_b$ in the hard kernel of BBNS
approach. We neglect the mass difference of b quark  and the B
meson. The scale $\mu$ in $\alpha_s$ is the renormalization scale
which is chosen as $\mu\sim {\cal O}(m_b)$ to eliminate the large
logarithms in the loop calculation. This scale is also the
factorization scale which separates the long- and short-distance
dynamics. The contribution from the momentum larger than $m_b$ is
involved in the hard scattering kernel while the contribution
from the momentum lower than $m_b$ is contained in the light-cone
distribution amplitude. As we will see, in the modified PQCD
approach, the scales become rich.

Before the discussion of the modified PQCD approach, it is
necessary to consider the most general case of PQCD method in
exclusive process. According to \cite{Brodsky}, a physical
quantity $M$ is given in terms of the hadronic wave function and
the hard scattering kernel in general:
\beq M=\int[dx][d^2 {\bf
k_T}]\prod_i\psi_i(x_i, Q, k_{Ti}) T(x, Q, k_T)
\eeq
where $Q\gg \Lambda_{QCD}$ is the large scale involved
in a process. If the transverse momentum $k_T$ can be negligible
in $T(x, Q, k_T)$, the above
formula can be simplified as
\beq M=\int[dx]\prod
_i\phi_i(x_i)T(x, Q) \eeq
In the above equation, we have used the
relation
\beq \phi(x, Q)=\int d^2{\bf k_T}\psi(x, Q, k_T) \eeq
The BBNS approach is just the application of the formula
(3) in B decays under the assumption that the form factor
is soft dominated.

If the end-point singularity can not be removed in the
convolution, the transverse
momentum in hard kernel cannot be neglected. As we
discussed in Sect. 1, a reliable treatment of the transverse
momentum effects must consider the mechanism of Sudakov
suppression. Once the effects of transverse momentum and
Sudakov form factor are taken into account, a transverse b-space
factorization formula \cite{Sterman2} will be obtained. In B
decays, this modified PQCD factorization formula is the
convolution of both the longitudinal momentum fraction  and the
transverse impact parameter $b$: \beq M(B\to \pi_1 \pi_2)= & \int
[d\xi][d^2 {\bf b}]{\cal P}_B(\xi, Q, b_1, \mu){\cal
P}_{\pi_1}(u, Q, b_2, \mu)  \\\nonumber &  \cdot {\cal
P}_{\pi_2}(v, Q, b_3, \mu)T(\xi, u, v, Q, b_1,b_2, b_3) \eeq
where $b$ is the conjugate variable of the transverse momentum
$k_T$ and $[d\xi]=d\xi dudv$, $[{\bf d^2 b}]={\bf d^2 b_1}{\bf
d^2 b_2}{\bf d^2 b_3}$.

In Eq.(5), the function ${\cal P}(x, Q, b)$ is:
\beq {\cal P}(x, Q,
b) &=& \int d^2{\bf k_T} e^{-i{\bf k_T\cdot b} }\psi(x, Q, k_T)\\
\nonumber &=& e^{-[s(x, Q, b)+s(\bar x, Q, b)]}\phi(x,
\frac{1}{b}) \eeq It includes all leading logarithmic enhancement
at large $b$ which has been included in Sudakov form factor. For
the light meson, Sudakov form factor suppresses the large b
contribution, so it selects component of the light meson wave
function with small spatial extent. Thus, $\phi(x,
\frac{1}{b})\approx \phi(x,Q)$.

The hard kernel $T(\xi, Q, b)$ is the Fourier transformation of
the hard scattering kernel defined in momentum space \beq T(\xi,
Q, b)=\int [d^2 {\bf k_T}]e^{-i{\bf k_T}\cdot{\bf b}}T(\xi, Q,
{\bf k_T}) \eeq

The evolution of the function ${\cal P}$ satisfies \beq
\mu\frac{d}{d\mu}{\cal P}(x, Q, b, \mu)=-2\gamma_q{\cal P}(x, Q,
b, \mu) \eeq where $\gamma_q$ is the quark anomalous dimension in
axial gauge.

Unlike the BBNS approach, there are many scales in the modified
PQCD approach, such as $uvQ^2$, $\frac{1}{b}$, etc.
In the modified PQCD approach, the scale parameter $\mu$ should
take the largest value of them. The
renormalization group equation must be used to eliminate the large
logarithm between many scales.

From the above discussion, we may expect that the BBNS approach and
the modified PQCD approach should be equivalent if the hard
kernel $T$ is fully hard dominated. It is really so. For the
case that the scattering kernel $T$ in Eq.(5) is concentrated
near $b\sim \frac{1}{Q}$, the Sudakov form factor is unity
and the function ${\cal P}(x, Q, b)$ is replaced by
distribution amplitude $\phi(x, Q)$. The modified PQCD
formula of Eq.(5) will be reduceded
into the BBNS factorization formula. If  the contributions of
${\cal O}(k_T^2\sim \Lambda_{QCD} M_B)$ in the scattering kernel
$T(x, Q, k_T)$ in Eq.(5)  are important so that
the end-point singularity of
$T(x, Q, k_T)$ at $k_T=0$ can not be removed in the
convolution, the two approach
will be different. In \cite{Sterman2}, an intuitive argument hold
that summed to all orders, the two approach are equivalent at
leading power in $1/Q^2$. However, at finite order, their
difference is unavoidable. The proof of BBNS approach needs the
heavy quark limit. While the modified approach enlarges the range
of PQCD down to accessible energies with the help of Sudakov
suppression.

\subsection*{3. Sudakov double logarithm and resummation}

Sudakov form factor comes from the
summation of the double logarithms to
all orders. In QED, the vertex correction in Feyman gauge gives
rise to Sudakov double logarithm ${\rm ln}^2 \frac{Q^2}{m_e^2}$
where $Q$ is the large energy scale and $m_e$ the electron mass.
The summation of the Sudakov double logarithms to all orders is
the exponential of the one-loop result. However, the non-abelian
theory QCD is more complicated than QED.
First, there is gluon self interaction in QCD which
makes the coupling constant large at low energy, so it
is necessary to consider the next-to-leading-log approximation.
Second, the light quark mass is smaller than the QCD scale
$\Lambda_{QCD}$, it cannot be taken as the infrared regulator.
The technic to perform the summation of double logarithms to all
orders in QCD is called resummation. In this case, it is Sudakov
resummation \cite{Collins,Sterman1}. Although the
resummation technic is fruitful and has been known for more than
ten years, its intricacy makes it difficult to understand. To
apply the resummation into a new process is more difficult. So,
it is necessary to discuss the main idea of the Sudakov form
factor without involving the intricate technic.

The Sudakov double logarithm is produced through the
overlap of collinear and soft divergence. The transverse momentum
degree is used to regulate the infrared divergence. The
calculation is performed in the transverse configuration
$b$ space instead of the momentum $k_T$ space.
The advantage of using the transverse $b$ space
is analyzed in \cite{bspace}. The momentum conservation
is automatically maintained in $b$ space and it is not necessary
to make any further assumptions about the transverse momentum
$k_T$ in higher orders. Moreover, in momentum $k_T$ space, it is
difficult to perform the next-to-leading-log approximation.

The collinear divergence in the massless limit depends on the
choice of gauge. In axial gauge, or
say physical gauge $n\cdot A=0$, the gluon propagator $D_{\mu\nu}$
satisfies $n^{\mu}D_{\mu\nu}=0$. So the non-factorizable collinear
divergence diminishes in axial gauge. This simplifies the
analysis of  factorization and Sudakov form factor. We will choose
the axial gauge in this section for discussion. However, the
obtained Sudakov form factor is gauge independent. A recent
literature about this conclusion can be found in \cite{Li2}.

In axial gauge, the Sudakov double logarithm occurs only in the
two-particle reducible diagrams. Thus, the Sudakov form factor is
included in each wave function itself, i.e., it is universal,
process-independent. The double logarithms at ${\cal O}(\alpha_s)$
are given by \cite{Sterman1}
\beq
I & =&
-\frac{C_F}{4\pi^3}\int_{_{l_T<Q}}\frac{d^2 {\bf
l_T}}{l^2_T}g_s^2(l_T)(e^{i{\bf l_T\cdot
b}}-1)\int^Q_{l_T}\frac{dl^+}{l^+}
\\\nonumber
& \approx & -\frac{2C_F}{\beta_1}{\rm ln}
(\frac{Q^2}{\Lambda^2}_{QCD})
{\rm ln}[\frac{{\rm ln}(\frac{Q^2}{\Lambda^2_{QCD}})}{{\rm
ln}(\frac{1}{b^2\Lambda^2_{QCD}})}]
\eeq
In the above equation,
we have chosen the light-cone variable and
$\beta_1=\frac{33-2n_f}{12}$. The factor of $e^{i{\bf l_T\cdot
b}}$ comes from the Fourier transformation from the transverse
momentum space to b-space. The occurrence of double logarithm
requires two condition: (1) two scales, $Q\gg\frac{1}{b}\gg
\Lambda_{QCD}$; (2) the overlap of collinear and soft regions. In
Eq.(9), the lower limit of $l^+$ must be in the soft region. To
sum the leading and next-to-leading logarithms to all order, it
needs to solve the renormalization group equation below
\cite{Collins, Sterman1}:
\beq Q\frac{\partial}{\partial Q}{\cal P}(x, Q,b)=
[K(b\mu)+\frac{1}{2}G(\frac{xQ}{\mu})+\frac{1}{2}G(\bar x
\frac{Q}{\mu})]{\cal P}(x, Q, b)
\eeq
where the functions of $K$
and $G$ satisfies
\beq \mu\frac{d}{d\mu}K=-\gamma_K , ~~~~~~~~
\mu\frac{d}{d\mu}G=~\gamma_K
\eeq
where $\gamma_K$ is anomalous
dimension. The functions $K$ and $G$ only depend on one scale:
$K$ is independent of large scale $Q$ and $G$ is independent of
scale $\frac{1}{b}$. The scale $\mu$ in $K$ and $G$ takes
different value: $\mu\sim {\cal O}(\frac{1}{b})$ in $K$;  $\mu\sim
{\cal O}(Q)$ in $G$. The appearance  of different scales can be
compared with the one scale case in BBNS approach. Solve the above
differential equations, one will obtain a Sudakov form factor in
distribution function \beq {\cal P}(x, Q, b)=e^{-[s(x, Q,
b)+s(\bar x, Q, b)]}\phi(x, \frac{1}{b})\eeq The definition of
${\cal P}$ function is shown in Eq.(6).

The Sudakov form factor $e^{-s}$ falls off quickly in large
$b$, or soft region and vanishes as $b>1/\Lambda_{QCD}$. Therefore
it suppresses the long-distance contribution, which is called
Sudakov suppresion. The behavior of Sudakov form facttor with
the vriable $b$ is plotted in Figure.\ref{figsuda}. The physical
reason is that an isolated colored parton tends to
radiate gulons. As $b$ increases, the color dipole associated
with quark and antiquark becomes more isolated, and they would
have more tendency to radiate gluons. In exclusive process,
however, the gluon radiation is forbidden by definition. So
the process with large $b$ separation will be suppressed.
Sudakov form factor manifests this phenomena in theory.
For small $b$, Sudakov form factor provides no suppression,
this region is dominated by hard scattering. In summary,
the Sudakov effects make small $b$ contributions dominant.
By including the Sudakov effects the
effective scale of the subprocess is $O(\Lambda_{QCD}Q)$.
As we have discussed, the Sudakov form factor is
universal. This simplifies the application of this effects  in
exclusive processes.

\subsection*{4. Sudakov effects in BBNS approach }

In $B\to \pi\pi$ decays, the two light pions carry the energy
of $\frac{m_B}{2}$ and moves fast away from the decay point. In
\cite{Beneke}, the authors argue that for realistic $b$ quark
mass, Sudakov from factor is not sufficiently effective. As
discussed in the last section, the Sudakov form factor is a
perturbative result. It plays a role in presence of the large scales
$Q\gg k_T\gg\Lambda_{QCD}$. The fact that the Heavy Quark
Effective Theory works very well and some successful
prediction of PQCD in
inclusive B decays implies that $m_b$ scale is large enough to
ensure the perturbative analysis. Moreover, BBNS approach
underlies the assumption of the heavy quark limit. In this limit,
the effectiveness of the perturbative Sudakov form factor is
obvious.

\begin{figure}[htbp]
   \begin{center}
      \epsfig{file=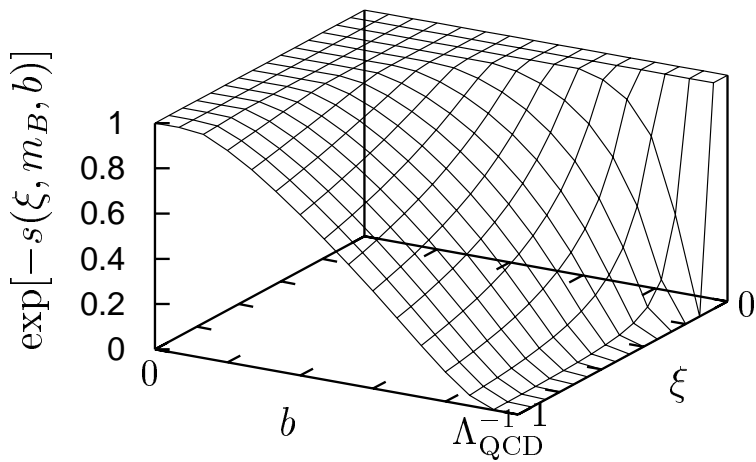, width=10cm}
   \vspace{-1.5cm}
  \end{center}
 \vspace{0.7cm}
 \caption{$b$-$\xi$ dependence of $e^{-s}$.}
\label{figsuda}
\end{figure}

For light pion meson, the Sudakov form factor is known. Its
explicit form can be found in \cite{Sterman1,Li3}. For simplicity,
we do not present it here again. For the
heavy meson, such as B meson, the heavy quark carries the most
energy while the light quark carries the momentum about
$\Lambda_{QCD}$. The wave function of B meson is soft dominated.
For $b$ quark, there is no collinear divergence thus the Sudakov
form factor is absent for it. For the light quark in B meson, its
longitudinal momentum mostly lies in the soft region. It seems
that there is no overlap of the collinear and soft regions. In
general, the soft dominance does not exclude the possibility that
the light quark may have the large longitudinal momentum although
this possibility is very small. For the case that the
longitudinal momentum of the light quark in B meson is small,
i.e., $\xi$ is small, Sudakov form factor contributed by the light
quark is $e^{-s(\xi,m_B,b)}\simeq 1$,(see Figure. \ref{figsuda}),
here $\xi$ is the momentum fraction of the light quark. Because
the large possibility is that the light quark of B meson only
carries small momentum which is around the order of
$\Lambda_{QCD}$, $\xi$ is dominantly distributed in the small
region around $\Lambda_{QCD}/m_B$. The possibility of large
$\xi$ is seriously suppressed by the B meson wave function.
Thus the Sudakov form factor for B meson only gives small
effect (Mostly it appraximately equals to 1). In this paper,
we give a Sudakov
form factor for the light quark in B meson for general
consideration as in \cite{Li1}. The numerical results in our study
show that the difference between the cases with and without Sudakov
form factor for B meson is less than $10^{-2}$ because
of the soft dominance of  B meson wave function.

\begin{figure}[htbp]
   \begin{center}
      \epsfig{file=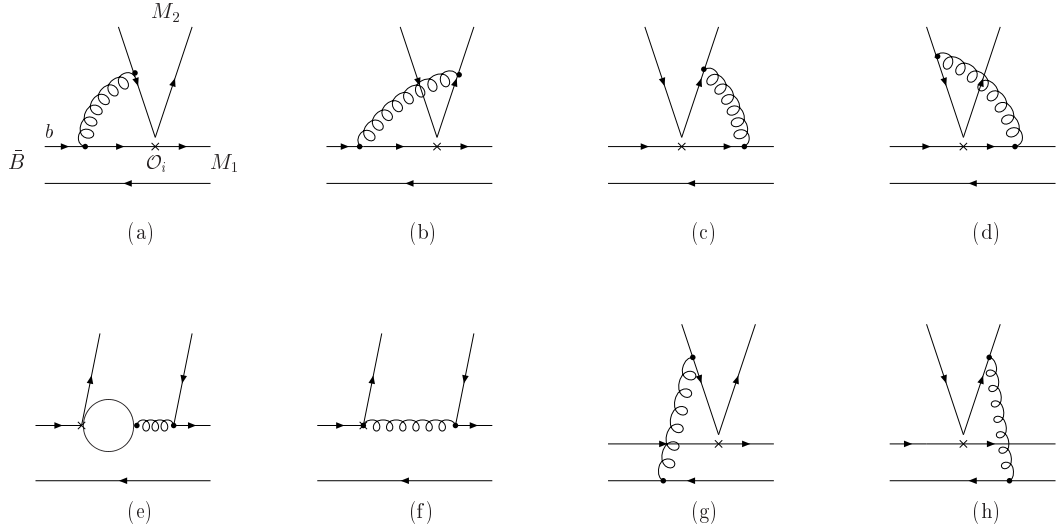, width=14cm}
   \vspace{-1.5cm}
  \end{center}
 \vspace{0.7cm}
 \caption{Order $\alpha_s$ corrections to the hard scattering kernels
 $T_{i}^I$ and $T_{i}^{II}$. (a)-(d): vertex corrections,
 (e) and (f): penguin correction, (g) and (h): hard spectator
 scattering. }
\label{feynman}
\end{figure}

Now it's time to discuss the Sudakov effects in $B\to \pi\pi$
decays. We restrict our discussion in the process analyzed in
\cite{Beneke}. The contributions to the $B\to \pi$ transition form
factor and the annihilation diagram are not discussed here. We
will give a detailed study about them in the next research. As in
\cite{Beneke}, we discuss the vertex correction, penguin
correction and the hard spectator scattering,
see Figure \ref{feynman}.

For the vertex corrections, the soft and collinear divergences
cancel for twist-2 \cite{Beneke} and symmetrical twist-3
distribution amplitudes \cite{Du} in the collinear limit. If
considering the transverse momentum effects, these
non-factorizable radiative corrections contribute subleading
logarithms in axial gauge. The leading contribution is double
logarithms which have been summed to a Sudakov form factor. The
subleading logarithms come from the soft gluon  where all
the four components of its momentum becomes soft.
This soft contributions cancel in
the collinear limit required by the factorization theorem.
As pointed out before, Sudakov form factor suppresses
the large $b$ region and
makes the dominate contribution come from the small $b$ region
which is near the collinear limit. In \cite{Sterman1}, the authors
study this soft gluon contribution with the transverse momentum
effects in hadron-hadron scattering. The non-factorizable soft
contributions are summed to all orders by using the renormalization
group equation. Their conclusion is that the soft contribution is
small and can be neglected. So the Sudakov effects in the vertex
corrections is expected to be small, and this conclusion is also
applicable for the penguin corrections.

The hard spectator scattering is depicted in
Figure.\ref{feynman} (g) and (h).
The partons in meson has the transverse momentum as well as the
longitudinal momentum. Compared to the collinear limit, the
mementum of partons in meson ( with momentum $P_1$ ) changes to
\beq k_1=uP_1+k_{T1},
~~~~~~k_2=\bar u P_1-k_{T1}
\eeq
where $u$ and $\bar u$ denote
the longitudinal momentum fraction. Our form is slightly
different from that in \cite{Beneke2}.
Our treatment corresponds to
the case that the meson is on-shell and the parton is
slightly off-shell. The off-shellness of the parton
is proportional to $k_T^2$.

For the hard spectator scattering, the contribution of the operator
$(S-P)\bigotimes(S+P)$ insertion vanishes in
the total hard scattering. The contribution of the
operator $(V-A)\bigotimes(V+A)$ insertion is equal to that of
$(V-A)\bigotimes(V-A)$. So it only needs to consider the
contribution of $(V-A)\bigotimes(V-A)$ operator insertion.

The twist-3 distribution amplitude contributes power correction.
But at the realistic $m_b$ energy scale, the power correction
parameter $r_{\chi}=\frac{2m_{\pi^-}^2}{m_b(mu+md)}\sim{\cal
O}(1)$ is not small. So the twist-3 contribution should be
considered in B decays. The twist-2 and twist-3 distribution
amplitudes are defined by
\beq
<\pi^-(P)|\bar{d_{\alpha}}(x)u_{\beta}(y)|0>=
\frac{if_{\pi}}{4N_c}\int_0^1  du
e^{i(uP\cdot x+\bar u P\cdot y)}[\gamma_5\not
P\phi_{\pi}(u)\\\nonumber +\mu_{\pi}\gamma_5\phi_P(u)
-\mu_{\pi}\sigma^{\mu\nu}\gamma_5 P_{\mu}(x-y)_{\nu}
\frac{\phi_{\sigma}(u)}{6}]_{\beta\alpha}
\eeq
where
$\mu_{\pi}=\frac{m_{\pi}^2}{(m_u+m_d)}$. ~$\phi_{\pi}$,
$\phi_P$ and $\phi_{\sigma}$ are the twist-2 and twis-3
distribution amplitudes, respectively. In the asymptotic limit,
$\phi_{\pi}(u)=6u\bar u$, $\phi_P(u)=1$ and
$\phi_{\sigma}(u)=6u\bar u$.

First we discuss the hard spectator scattering contribution in
BBNS approach. The formula is derived in $k_T$ space and remains
the transverse momentum at the beginning. About the coordinate
variable $(x-y)_{\nu}$ in Eq.(14), we make the transformation to
project it into the momentum space as adopted in \cite{Du}:
\beq
(x-y)_{\nu}e^{-i(x-y)\cdot P}=
  i\frac{\partial}{\partial P^{\nu}}e^{-i(x-y)\cdot P}
\eeq

With this projection, the hard spectator scattering
contribution in Figure. \ref{feynman} (g) and (h)
 is formulated in transverse
momentum space:
\beq S_{g+h}= &
\displaystyle\frac{-if_Bf_{\pi}^2}{4N_c^2}
g_s^2C_F\int[d\xi][d^2{\bf k_T}]~~~~~~~~~~~~~~~~
~~~~~~~~~~~~~~~~~~~~~~~~~~~~~~~~~~~\\\nonumber
& \cdot [\displaystyle\frac{um_B^4
\phi_B(\xi)\phi_{\pi}(u)\phi_{\pi}(v)}{[\xi
um_B^2+({\bf k_T }-{\bf k_{T1}})^2][-uvm_B^2+({\bf k_T}-{\bf
k_{T1}}+{\bf k_{T2}})^2]} ~~~~\\\nonumber &   +2\mu_{\pi} m_B^5
\displaystyle\frac{-u v
\phi_B(\xi)\frac{\phi_{\sigma}(u)}{6}\phi_{\pi}(v)}{[\xi
um_B^2+({\bf k_T }-{\bf k_{T1}}^2)][-uvm_B^2+({\bf k_T}-{\bf
k_{T1}}+{\bf k_{T2}})^2]^2} ] \eeq
We have assumed the momentum fraction $\xi$ in B meson is small
and the distribution amplitudes are symmetric. Neglecting the
transverse momentum in both the numerator and the denominator will
give a simplified formula for the hard spectator scattering:
\beq  S_{g+h}=\frac{if_Bf_{\pi}^2}{4N_c^2}g_s^2C_F
\int d\xi
dudv[\frac{\phi_B({\xi})\phi_{\pi}(u)\phi_{\pi}(v)}{\xi uv }
+\frac{2\mu_{\pi}}{m_B}\frac{\phi_B({\xi})
\frac{\phi_{\sigma}(u)}{6}\phi_{\pi}(v)}{\xi u^2 v} ] \eeq

This formula is consistent with the corresponding one
given in \cite{Du}.
The scale $\mu$ in $g_s$ is chosen as $m_b$. For twist-2
distribution amplitude,  there is no end-point singularity. When
$u \rightarrow 0$, the twist-3 contribution will lead to
end-point singularity. The physical reason is that the virtual
gluon approaches to the mass shell. This is a soft logarithmic
divergence. As discussed in Sect. 1, the occurrence of
end-point singularity is the result of neglecting the transverse
momentum in the denominator.

In the modified PQCD approach, the partonic transverse momentum
is retained without assuming  $k_T^2\ll \xi um_b^2, ~uvm_b^2$.
The final formula contains the convolutions of the longitudinal
momentum fraction and the transverse impact parameter $b$,
\beq
S_{g+h} & = & \frac{-if_Bf_{\pi}^2}{4N_c^2}g_s^2C_F\int d\xi dudv
\int bdb b_2db_2 \\\nn & & \cdot
\{~~ u m_B^4 {\cal P}_B(\xi,b){\cal P}_{\pi}(u,b)
{\cal P}_{\pi}(v,b_2)K_0(-i\sqrt{uv}m_B b_2)\\\nonumber &  &
~~\cdot [\theta(b_2-b)I_0(\sqrt{\xi
u}m_Bb)K_0(\sqrt{\xi u}m_B b_2)  +
\theta(b-b_2)I_0(\sqrt{\xi u}m_Bb_2)K_0(\sqrt{\xi u}m_B b)
]\\\nonumber & & -2uv\mu_{\pi}m_B^5 {\cal P}_B(\xi,b)
\frac{{\cal P}_{\sigma}(u,b)}{6}{\cal P}_{\pi}(v,b_2)
 \frac{b_2}{-2i\sqrt{uv}m_B}K_{-1}(-i\sqrt{uv}m_B b_2)\\\nonumber & & ~~
\cdot[\theta(b_2-b)I_0(\sqrt{\xi u}m_Bb)K_0(\sqrt{\xi u}m_B b_2)+
 \theta(b-b_2)I_0(\sqrt{\xi
u}m_Bb_2)K_0(\sqrt{\xi u}m_B b)  ] ~~\}
\eeq where $K_i$ and $I_i$
are modified Bessel functions and $i$ is its order.

This formula is more complicated than the result of BBNS
approach. One can check that the result in the right hand
side of Eq.(18) is finite, and there is no divergence in it.
The modified QCD formula is
self-consistent and contains no arbitrary
phenomenological parameter except for the input
distribution amplitudes.

In the numerical calculation, the distribution amplitutes of pions
for twist-2 and twist-3 are taken as their asymptotic limit.
The distribution amplitude for B meson is $\phi_B(x,b)=N_B
x^2(1-x)^2 exp(-\frac{M_B^2x^2}{2\omega_B^2}-
\frac{1}{2}(\omega_Bb)^2)$ where
$\omega_B=0.3GeV$, $N_B$ is the normalization constant. The QCD
scale $\Lambda_{QCD}=0.3GeV$,  and the other input parmeters are
taken as follows: $f_B=0.19GeV$, $f_{\pi}=0.13GeV$,
$F^{B\pi}_0(0)=0.3$, $m_B=5.27GeV$, $\mu_{\pi}=1.2GeV$.

In our numerical result, the twist-3 contribution is not
important in hard spectator scattering. So, there is only a little
improvement in numerical value compared with the the former
calculations in BBNS approach \cite{Beneke, Du, Muta}.
We will not present the full calculation of $B\to \pi\pi$ decays
because it is unnecessary. The comparison of the prediction for
the hard spectator scattering in both BBNS approach and the
modified QCD approach is presented below.

Define $f$ as the value of the hard spectator scattering
contribution divided by the lowest order result
$f\equiv \frac{S_{g+h}}{if_{\pi}F_0^{B\pi}m_B^2}=f_2+f_3$
where $f_2$, $f_3$ represent the contribution of twist-2 and
twist-3 terms. The numerical result (For (V-A) $\bigotimes$ (V-A)
operator insersion) is:

{\bf Twist-2:} In BBNS approach, $f_2=0.043$; In the modified PQCD
approach, $f_2=0.057+i0.0037$;

{\bf Twist-3:} In BBNS approach, $f_3$ cannot be calculated, it
is expressed in terms of phenomenological parameters $\rho_H$ and
$\phi_H$ \cite{Beneke2},
\beq f_3=\frac{\pi\alpha_s C_F f_Bf_{\pi}}{m_B^2 F_0^{B\pi}N_c^2}
     \int^1_0
     \frac{d\xi}{\xi}\phi_{B}(\xi)\int^1_0\frac{dx}{x}\phi_{\pi}(x)
     \frac{2\mu_{\pi}}{m_b}(1+\rho_H e^{i\phi_H})
     {\rm ln}\frac{m_B}{\Lambda_h},
\eeq
where $\rho_H\le 1$, $\Lambda_h=0.5GeV$;
 In the modified PQCD approach, $f_3=0.0166-i0.0144$; Compare
the two results, we can get $\rho_H=0.97$, which is within the
constraint of $\rho_H\le 1$, and the strong phase is very large,
it is $\phi_H=-83.7^\circ$.

\subsection*{5. Conclusions and discussions}

Through the exchange of the gluons, the partons in hadron carries
the transverse momentum ${\bf k_T}$. Its effects is important in
the end-point region. Neglecting it will lead to the end-point
singularity in BBNS approach. The problem of end-point
singularity can be reliably treated in the modified PQCD approach.
Retaining the partonic intrinsic transverse
momentum and with the help of
Sudakov form factor, the modified PQCD approach is a
self-consistent, model-independent framework. For the vertex
corrections and the penguin corrections, the end-point
singularity in the hard  kernel
is cancelled in the convolution
and Sudakov suppression gives little effect.
The separation of long- and short-distance
dynamics is good enough to ensure the validity of factorization.
BBNS approach provides a successful, easy-to-do framework for these
diagrams. For the hard spectator scattering,
if neglecting the partonic intrinsic transverse momentum
the end-point singularity in the hard scattering kernel can not be cancelled
in the twist-3 case, which implys that the contributions from the end-point
region are important and such ampiltude can not be analyzed at a fixed
order in PQCD. In this case one has to include the transverse momenta
of partons and
Sudakov form factor in order to proceed at a fixed order in PQCD.
Sudakov form factor can suppress the soft contribution and make the hard
contribution dominant. In this case,
Sudakov correction is important. Our numerical results show that
Sudakov correction is small at leading twist level and important
at twist-3 level. The twist-3 contribution in the hard spectator
scattering is non-negligible, but not dominant.

In \cite{Beneke}, it is argued that the transverse momentum
effect is power suppressed so that it can be neglected.
This is valid only in the $m_b\rightarrow \infty$ limit.
Acturally, in the loop corrections, the large logarithms
such as ${\rm ln}^2\frac{Q^2}{k_T^2}$, ${\rm ln}\frac{Q^2}{k_T^2}$
etc will occur. In the tree level, the hard kernel contains the
terms such as $\frac{1}{uvm_B^2+k_T^2}$,
$\frac{1}{(uvm_B^2+k_T^2)^2}$. Dropping the transverse momentum,
or set it to zero, will lead to and end-point
singularity which will destroy the factorization theorem. This is
the reason to incorporate the Sudakov effects. Including the
Sudakov effects, the naive power counting in \cite{Beneke} will
be modified. The contribution of the end-point region is smeared
by the transverse momentum effects.
So the assumption that the $B\to
\pi$ transition form factor is dominated by soft end-point
interaction may be questionable. It can be hard momentum
transfer doimnant.
Recently a complete PQCD method was applied to the study of
two-body $B$ meson decays of $B\to \pi\pi$, $K\pi$, and so on,
including completely perturbative treatment of $B\to\pi$,
$B\to K$ transition form factors and annihilation diagrams
\cite{PQCD}. Some interesting results have been obtained.
However, this approach has been upgrading continuously.
It seems that there is still a bit of long way
to go before getting final success.
A systematic analysis about
the $B\to \pi$ transition form factor, the annihilation diagram
and the radiative corrections is still needed. Except these
problems, another important subject is to understand the
factorization theorem in B decays. Up to now, some
works along this direction has been done \cite{Beneke, Li4}.
More detailed works on the proof of factorization theorem
in B decays to all orders is still highly needed.
Without the proof of factorization
theorem, any formulas can only be regarded as a ``model".
In one word, it is desirable to carefully consider the
factorization in B decays.

\section*{Acknowledgment}

Two of the authors (Z. Wei, M. Yang) would like to thank Prof.
Hsiang-Nan Li for his useful  discussion and his interesting
talks. Wei also expresses his gratitude to Prof. Hai-Yang Cheng
for his warmful discussions about the BBNS approach. This work is
supported in part by National Natural Science Foundation of China
and the Grant of State Commission of Science and Technology of
China. M. Yang thanks the partial support of the Research
Fund for Returned Overseas Chinese Scholars.


\begin{thebibliography}{99}
\bibitem{exc}F.  Farrar and D. Jackson,
    \Journal{\PRL}{43}{246}{1979};
    S.  Brodsky and G.  Lepage,
    \Journal{\PRL}{43}{545}{1979},
    \Journal{\PLB}{87}{359}{1979};
    A. Efremov and A. Radyushkin,
    \Journal{\PLB}{94}{245}{1980};
    A. Duncan, A. Mueller,
    \Journal{\PRD}{21}{1636}{1980}.

\bibitem{Brodsky} G. Lepage and S. Brodsky, Phys.Rev.D{\bf 22}
        (1980) 2157.

\bibitem{Beneke} M. Beneke, G. Buchalla, M. Neubert, C.T.
        Sachrajda, Phys.Rev.Lett.{\bf 83} (1999) 1914-1917;
        Nucl.Phys.B591 (2000) 313-418.

\bibitem{BSW} M. Bauer, B. Stech and M. Wirbel, Z.Phys.C {\bf 34},
        103 (1987); \\M. Wirbel, B. Stech and M. Bauer,
         Z.Phys.C {\bf 29}, 637 (1985).

\bibitem{Du} D. Du, D. Yang, G. Zhu,
        Phys.Lett.B{\bf 509} (2001) 263;
        Phys.Rev.D{\bf 64} (2001) 014036.

\bibitem{Beneke2} M. Beneke, G. Buchalla, M. Neubert, C.T.
        Sachrajda, hep-ph/0104110.

\bibitem{Sterman2} H. Li, G. Sterman, Nucl.Phys.B{\bf 381} (1992)
        129-140;  G. Sterman, P.Stoler, hep-ph/9708353.

\bibitem{Li1} H. Li, H. Yu, Phys.Rev.Lett.{\bf 74} (1995) 4388;
             T. Yeh, H. Li, Phys.Rev.D{\bf 56} (1997) 1615-1631.

\bibitem{Collins} J. Collins, D. Soper, Nucl.Phys.B{\bf 193}
        (1981) 381.

\bibitem{Sterman1} J. Botts, G. Sterman, Nucl.Phys.B{\bf 325} (1989)
        62.

\bibitem{bspace} Y. Dokshitzer, D. D'Yakonov, S. Troyan,
        Phys.Lett. B{\bf 488} (2000)

\bibitem{Li2} H. Li, Phys.Rev.D{\bf 55} (1997) 105.

\bibitem{Li3} H. Li, Phys.Rev.D{\bf 52} (1995) 3958.

\bibitem{Muta} T. Muta, A. Sugamoto, M. Yang, Y. Yang,
        Phys.Rev.D{\bf 62} (2000) 094020;
              D. Du, D. Yang, G. Zhu,
        Phys.Lett. B{\bf 488} (2000) 46.

\bibitem{PQCD}Y. Keum, H, Li, A. Sanda, Phys.Lett.B{\bf 504}(2001)6;
    C. Chen, H, Li, Phys.Rev.D{\bf 63} (2001) 014003;
    Y. Keum, H, Li, A. Sanda, Phys.Rev.D{\bf 63} (2001) 054008;
    Y. Keum, H. Li, Phys.Rev.D{\bf 63} (2001) 074006;
    C. Lu, K. Ukai, M, Yang, Phys.Rev.D{\bf 63} (2001) 074009.

\bibitem{Li4}H. Li, Phys.Rev.D{\bf 64} (2001) 014019;~~ C. Bauer,
    D. Pirjol, I. Stewart, hep-ph/0107002.

\end{thebibliography}
\end{document}